\documentstyle[aps,prb,multicol,epsf]{revtex}
\begin{document}

\title{Transport in Materials with Disclination Dipoles:\\
Applications to Polycrystals and Amorphous Dielectrics}

\author{S.E. Krasavin and V.A. Osipov}
\address{
Joint Institute for Nuclear Research,\\
Bogoliubov Laboratory of Theoretical Physics\\
141980 Dubna, Moscow region, Russia}

\address{\em (\today)}
\preprint
\draft
\maketitle

\begin{abstract}
The problem of both electron and phonon scattering by wedge
disclination dipoles (WDD) is studied in the framework of the
deformation potential approach. The exact analytical results for
the mean free path are obtained within the Born approximation. The
WDD-induced contribution to the residual resistivity in
nanocrystalline metals is estimated. Using the WDD-based model of
a grain boundary, the thermal conductivity, $\kappa$, of
polycrystals and amorphous dielectrics is studied. It is shown
that the low-temperature crossover of $\kappa$ experimentally
observed in LiF, NaCl, and sapphire can be explained by the
grain-boundary phonon scattering. A combination of two scattering
processes, the phonon scattering due to biaxial WDD and the
Rayleigh-type scattering, is suggested to be of importance in
amorphous dielectrics. Our results are in a good agreement with
the experimentally observed $\kappa$ in a-SiO$_2$, a-GeO$_2$ a-Se,
and polystyrene over a wide temperature range.
\end{abstract}
\vskip 0.2cm

\begin{multicols}{2}

\section{Introduction}

There exist many varieties of extended defects in crystals,
topological in their origin. The most known examples are
dislocations, disclinations, twins, grain boundaries, stacking
faults etc. These defects play a significant role in description
of various phenomena in real crystals as well as in
non-crystalline materials. In particular, there is reason to
believe that linear defects like dislocations and disclinations
are the principal imperfections of liquid crystals~\cite{kleman},
some amorphous solids~\cite{nelson,rivier}, and
polymers~\cite{li}.

The contribution to the transport characteristics due to
dislocations is now well understood (see, e.g.,
Refs.~\cite{ziman,gantmakh}). Some aspects of the qualitative
behavior of the disclination-induced electron scattering have been
recently presented~\cite{pla97}. In particular, it was found that
both dislocations and disclinations can be  effective scattering
centers for conducting electrons, especially at low temperatures
when other scattering mechanisms are suppressed. Thus, along with
point impurities, these defects give a contribution to the
residual resistivity. In real crystals, however, the isolated
disclinations are rather exotic objects. Instead, for topological
reasons, the dipole configurations are more favourable. In
addition, the creation energy for a single disclination
considerably exceeds that for a disclination dipole.

There is another reason to call attention to dipoles of
disclinations. As a matter of fact, the wedge disclination dipoles
(WDD) simulate finite dislocation walls. In turn, dislocation
walls describe the low-angle grain
boundaries~\cite{read,amelinckx}. Thus, one can  expect that the
results obtained for disclination dipoles will be useful in
description of the grain-boundary scattering problem. This allows
us to extend the possible applications to a study of the transport
properties of polycrystals where grain boundaries are of
importance. In particular, there is experimental evidence that the
grain boundaries give contribution to the resistivity in metals
(see, e.g., review in Ref.~\cite{gusev}) which in part depends on
the size of the grain boundary. Notice that though the problem of
the grain-boundary-induced scattering has been formulated many
years ago the proper solution is still absent.

A theory of the phonon scattering by grain boundaries has been
developed within the Born approximation in Ref.~\cite{klemens}. A
grain boundary was considered as a wall of edge dislocations with
a rather strong assumption that the dislocation wall is {\it
infinitely} long. Nevertheless, the main properties predicted by
this model have been confirmed by
experiments~\cite{berman,anderson,roth}. In particular, the phonon
mean free path was found to be a constant over a wide temperature
range. As a result, the low-temperature thermal conductivity in
polycrystals varies as $T^3$ in agreement with the experimental
data. However, the model of an infinitely long dislocation wall
failed to describe a remarkable increase in thermal conductivity
observed below some characteristic temperature $T^{'}$ ($T^{'}\sim
0.1$K for LiF~\cite{anderson,roth}).

Recently~\cite{krasavin1}, the grain boundary phonon
scattering problem has been investigated within the more realistic
model which takes into account the finiteness of the boundary. The
basis for this model was the known analogy between disclination
dipoles and {\it finite} walls of edge
dislocations~\cite{li1,deWit}. It was found that the proper
consideration of the finiteness of the boundary results in the
low-temperature crossover of the thermal conductivity in agreement
with experiments~\cite{anderson,roth}.

A more intriguing application is an attempt to describe the
physics of amorphous dielectrics by considering WDD as the basic
structural elements~\cite{krasavin2}. This point of view agrees
with the cluster picture proposed earlier for
glasses~\cite{kauzm,phill}. In accordance with this picture there
exist many small crystalline grains (microclusters with average
diameters of order 20-30\AA) in the amorphous state. On the other
hand, the concept of elastic dipoles for orientational glasses was
introduced and successfully explored in
Refs.~\cite{randeria,sethna1}. The authors considered elastic
dipoles as the additional to two-level systems (TLS) structural
elements of glasses. This assumption allows them to describe both
the specific heat and the thermal conductivity of some glasses in
the plateau region and above (up to $100$K). The physical nature
of these dipoles, however, has not yet been clarified. As is
known~\cite{ziman}, a possible way to understand the
microstructure as well as the nature of principal imperfections is
to study the transport properties of a material. It will be shown
below that the concept of the elastic dipole can be successfully
realized via dipoles of wedge disclinations without considering
TLS. Namely, the WDD-induced phonon scattering is found to provide
the correct description of the low-temperature thermal
conductivity of amorphous dielectrics. Notice also that in
accordance with the geometrical consideration disclinations are
expected to always be present in the amorphous
state~\cite{kleman1} (see also Ref.~\cite{nelson}).

The main goal of the paper is twofold. First, we outline the
general formalism of the WDD-induced scattering for both electrons
and phonons. The most important details needed for a better
understanding of calculations are given. Second, we apply the
results obtained for description of two important problems. The
first one is the experimentally observed deviation of the thermal
conductivity from a $T^3$ dependence below 0.1 K in LiF and NaCl.
The second problem is the thermal conductivity of amorphous
dielectrics.
It was shown in Ref.~\cite{krasavin2} that the experimentally
observed behavior of thermal conductivity of a-SiO$_2$ over a wide
temperature range can be explained by a combination of two
scattering processes. The first one comes from the phonon
scattering due to biaxial WDD while the second one is the Rayleigh
type scattering. In the present paper we extend our investigation
to other glasses and discuss some unsolved questions.

The paper is organized as follows. The general formalism of the
WDD-induced scattering is presented in Sec.II. We consider all
possible types of WDD and calculate the corresponding phonon mean
free paths. The principal distinction between the scattering
properties of the uniaxial and biaxial dipoles is shown. The
developed approach is applied to estimation of the WDD-induced
contribution to the residual resistivity of granular metals in
subsection A. The phonon scattering due to WDD is studied and the
contribution to the thermal conductivity is calculated in
subsection B. The obtained results are compared with the
experimental data for LiF. In Sec.III we apply the WDD-induced
mechanism of phonon scattering to the problem of thermal
conductivity in amorphous dielectrics. The results are compared
with the experimentally observed $\kappa$ in a-SiO$_2$, a-GeO$_2$
a-Se, and polystyrene (PS). Sec.IV is devoted to the detailed
discussion of the results obtained, specifically with relation to
the proposed WDD-based model for dielectric glasses.

\section{Theory of the WDD-induced scattering}

In this section we study the contribution to the effective
cross-section which comes from the potential associated with a
static deformation of a lattice caused by straight WDD. Two
reasonable approximations are in common usage in studies of such
problems (see, e.g., Refs.~\cite{ziman,gantmakh}). First, we
suppose that the scattering processes are elastic and, second, the
Born approximation is valid. Besides, we will consider here the
simplest case when the only elastic deformations are dilatations.

In this case, an effective perturbation energy
due to the strain field caused by a single WDD takes the form
\begin{equation}
\label{eq1}
U({\bf r}) =GSpE_{ij},
\end{equation}
where $E_{ij}$ is the strain tensor and $G$ is an interaction
constant.

Let the disclination lines be oriented along the z-axis, the
position of the positive disclination be (0,-L) while of the
negative one (0,L) (see Fig.1).

\vskip 0.5cm
\begin{figure}[htb]
\epsfxsize=7.0cm \centerline{\hspace{0mm} \epsffile{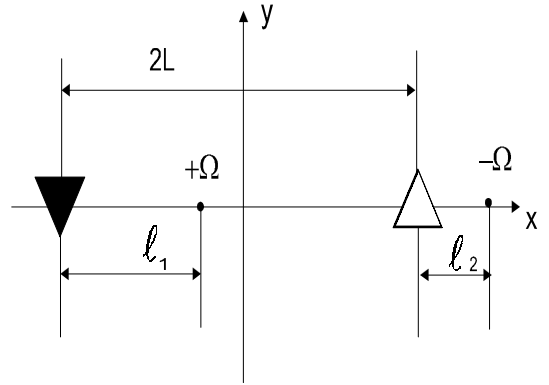}}
\vspace{3mm}
\caption{Schematic picture of a wedge disclination
dipole with shifted by distances $l_{1}$ and $l_{2}$ axes of
rotation. The disclination lines are oriented along the $z$-axis
for both defects}
\label{fig1}
\end{figure}

 Notice that in the general case the axes of rotations (with ${\bf \Omega _{1}}=\Omega {\bf e_{z}}$
and ${\bf \Omega _{2}}=-\Omega {\bf e_{z}}$) are shifted relative
to their positions by arbitrary distances $l_{1}$ and $l_{2}$,
respectively. For $l_{1}=l_{2}=0$ one gets the biaxial WDD with
nonskew axes of rotation. It was shown~\cite{li2} that this dipole
can be simulated by a finite wall of edge dislocations with
parallel Burgers vectors. In particular, the far strain fields
caused by biaxial WDD agree with those from a finite dislocation
wall~\cite{deWit}.

For $l_{1}-l_{2}=2L$ and $l_{1}=-l_{2}$ one gets the uniaxial and
the symmetrical uniaxial WDD, respectively. Notice that uniaxial
WDD can be simulated by a finite wall of edge dislocations
complemented by two additional edge dislocations at both ends of
the wall. The sign of these dislocations is opposite to that of
the dislocations in the wall and absolute values of Burgers
vectors are equal to $b=2L\tan(\Omega/2)$ ($b=b_{y}$ at chosen
geometry). As a result, the uniaxial WDD becomes a strongly
screened system as opposed to the biaxial WDD (see Fig.2). In the
general case, $l_1\not=l_2\not=0$, one gets the biaxial WDD with
shifted axes of rotation.

\vskip 1.5cm
\begin{figure}[h]
\epsfxsize=7.5cm \centerline{\hspace{0mm} \epsffile{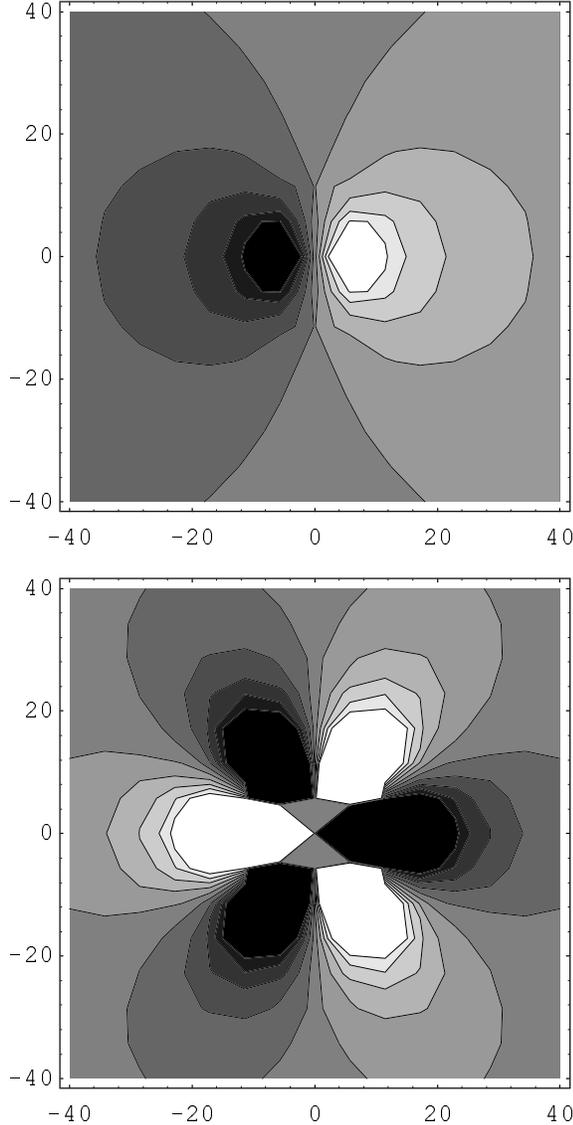}}
\vspace{1.5mm}
\caption{ Contour plot of the perturbation energy (2)
$U(x,y)/B$ in arbitrary units for biaxial WDD ($l_{1}=l_{2}=0$,
upper picture) and for uniaxial WDD ($l_{1}-l_{2}=2L$, lower
picture). } \label{fig2}
\end{figure}
Substituting the explicit form of $E_{ij}$ into Eq.(1) (see Apendix A) we
find for the perturbation energy $U({\bf r})$
$$
\label{eq2}
U(x,y)=B\Biggl [\frac{1}{2}\ln {\frac{(x+L)^2+y^2}{(x-L)^2+y^2}}-
l_1\frac{x+L}{(x+L)^2+y^2}
$$
\begin{eqnarray}
+l_2\frac{x-L}{(x-L)^2+y^2} \Biggr ],
\end{eqnarray}
where $B=G\nu(1-2\sigma)/(1-\sigma)$,
$\nu=\Omega/2\pi$ is the Frank index, and $\sigma$ is the Poisson constant.
Notice that all possible types of WDD are included in Eq.(2).

As is seen from Eqs.(A1) and (A2) in Appendix, all strains caused
by WDD are located in the $xy$-plane. In this case, the only
components of the wavevector that are normal to the defect lines,
${\bf q}={\bf q_{\bot}}$, are involved in the scattering process.
For the sake of simplicity, let us assume that incident carriers
are normal to disclination lines.

The problem reduces to the two-dimensional case with the mean free path
given by
\begin{equation}
\label{eq3}
\Lambda^{-1} = n_{i}\int_0^{2\pi}(1-\cos\theta)\Re(\theta)d\theta.
\end{equation}
Here $\Re(\theta)$ is an effective differential scattering radius,
and $n_i$ is the areal density of WDD. Notice that when axes of
WDD are oriented randomly, one has to perform an additional
averaging over all possible angles of incidents. As was shown for
dislocations~\cite{gantmakh}, however, such averaging leads merely
to a modification of the numerical factor in $\Re(\theta)$.

Within the Born approximation, $\Re(\theta)$ is determined to
be~\cite{ziman}
\begin{equation}
\label{eq4}
\Re(\theta) = \frac{qS^2}{2\pi\hbar^2v^2}
\overline{\left|<{\bf q}|U({\bf r})\right|{\bf q^{'}}>|^2},
\end{equation}
where all vectors are two-dimensional, $S$ is a projected area,
${\bf v}$ is a carrier velocity,
the bar denotes an averaging procedure over
$\alpha$ which defines an angle between ${\bf p}={\bf q} - {\bf q^{'}}$
and the $x$-axis. In other words, it means the averaging over randomly
oriented dipoles in the $xy$ plane.
Evidently, the problem reduces to the estimation of
the matrix element in Eq.(4) with the potential from Eq.(2).
For this purpose, it is convenient to use the polar
coordinates $(r,\phi)$
\begin{eqnarray}
\label{eq5}
U(p, \alpha) &=& <{\bf q}|U({\bf r})|{\bf q^{'}}> \nonumber \\
&=&\frac{1}{S}\int d^2{\bf r} \exp[ipr\cos(\phi-\alpha)]U(r,\phi).
\end{eqnarray}

For elastic scattering, the matrix element in Eq.(5) depends only on
$|{\bf q}|=|{\bf q^{'}}|$ and the scattering angle $\theta $. Thus,
$p=|{\bf p}|=|{\bf q}-{\bf q^{'}}|=2q\sin(\theta/2)$.

After integration in Eq.(5) and following averaging of $|U(p,\alpha)|^2$
with respect to $\alpha$ one obtains (see details in Appendix B)
\newpage
\begin{eqnarray}
\Re(\theta )=\frac{\pi B^2}{\hbar^2
v^2\sin(\theta/2)}\Biggl\{\frac{2}{p^3} \Bigl
(1-J_{0}(2pL) \Bigr)-\frac{2\Delta_l}{p^2}J_{1}(2pL)\nonumber \\
+\frac{\Delta^2_l}{2p}\Bigl(\frac{1}{2}+J_{0}(2pL))
-\frac{J_{1}(2pL)}{2pL}\Bigr)\Biggr\},
\end{eqnarray}
where $\Delta_{l}=l_{1}-l_{2}$, $J_{n}(t)$ are the Bessel functions.
Upon integrating Eq.(3) with respect to $\theta$ one finally obtains
$$
\label{eq7}
\Lambda^{-1}=\frac{B^2L^2\pi^2n_{i}}{q\hbar ^2v^2} \Biggl \{z^2\left (\frac{1}{2}+J_0^2(2qL)\right)
-\frac{4}{qL}J_0(2qL)J_1(2qL)
$$
\begin{eqnarray}
+\left (8-\frac{z(z+8)}{2}\right )\Bigl (J_0^2(2qL)+J_1^2(2qL)\Bigr)\Biggr \},
\end{eqnarray}
where $z=\Delta_l/L$. It should be emphasized that Eq.(7) is the
exact result which allows us to describe all types of WDD. Notice,
that the behavior of $\Lambda$ in Eq.(7) is actually governed by
the only parameter $2L$.

Let us consider two important limiting cases. For biaxial dipoles with
$\Delta_{l}=0$ ($z=0$) one obtains
\begin{eqnarray}
\Lambda_{bi}^{-1}=\frac{8B^2L^2n_{i}\pi^2}{q\hbar ^2v^2}
\Biggl \{J^2_{0}(2qL)+J^2_{1}(2qL)\nonumber \\
-\frac{1}{2qL}J_{0}(2qL)J_{1}(2qL) \Biggr \}.
\end{eqnarray}
For uniaxial dipoles $\Delta_l=2L$ $(z=2)$, and the mean free path
$\Lambda$ is
\begin{eqnarray}
\Lambda_{uni}^{-1}=\frac{2B^2L^2n_{i}\pi^2}{q\hbar ^2v^2}
\Biggl \{1+J^2_{0}(2qL)-J^2_{1}(2qL)\nonumber \\
-\frac{2}{qL}J_{0}(2qL)J_{1}(2qL) \Biggr \}.
\end{eqnarray}
In what follows, we apply the developed formalism to the problem
of the WDD-induced electron and phonon scattering.

\subsection{Electron scattering: WDD-induced residual resistivity in metals}

As is known~\cite{ziman}, the residual resistivity of metals may
be caused by electron scattering due to linear defects like
dislocations, stacking faults and grain boundaries. It was
mentioned in the introduction that the WDD-based model is a good
candidate for modelling the grain boundaries. Thus, the previous
analysis allows us to study the contribution to the residual
resistivity due to grain boundaries.

Let us use the well-known Drude formula for the resistivity
in the static regime
\begin{equation}
\rho=(\frac{m}{ne^{2}})\langle \frac{1}{\tau} \rangle ,
\end{equation}
where $\tau$ is the relaxation time, $m$ and $e$ are the mass and
the charge of the conducting electron, and $n$ is the electron
density. For point impurities and linear defects like dislocations
and disclinations, the angle brackets denote the configurational
average. In our case, this is averaging over $\alpha$ in Eq.(4).
Thus, one can write the final result
\begin{equation}
\rho=(\frac{mv_F}{ne^{2}})\Lambda^{-1}
\end{equation}
with $\Lambda$ from Eq.(7), where $q=q_F$, $v=v_F$, and $G=G_d$.
Obviously, index $F$ denotes the Fermi values, and $G_d\simeq
(2/3)E_F$ is the deformation potential constant~\cite{ziman},
where $E_F$ is the Fermi energy. For simplicity, we restrict our
consideration to metals of zinkblende or wurtzite structures. In
metals typically $q_{F}\approx 0.6\div 1.2$\AA$^{-1}\ $. A dipole
separation is chosen to be $2L\approx 10^2\div 10^4$\AA\  which is
of order of the grain size in polycrystals. In this case,
$2q_{F}L>>1$ and, consequently, the Bessel functions in Eq.(7) can
be approximated by their asymptotic values. The results are the
following
\begin{equation}
\rho_{bi}=\frac{16B_e^2\pi Ln_{i}}{ne^{2}m v^{3}_F},
\end{equation}
and
\begin{equation}
\rho_{uni}=\frac{2B_e^2\pi^2L^2n_{i}}{ne^{2}\hbar v^{2}_{F}},
\end{equation}
where $B_e=B(G=G_d)$.

The main difference in the behavior of $\rho_{bi}$ and
$\rho_{uni}$ comes from their $L$-dependence. As is seen, at fixed
$n_i$ $\rho_{bi}\sim L$ while $\rho_{uni}\sim L^2$. As the result,
$\rho_{bi}$ is found to be larger than $\rho_{uni}$. This agrees
with the above-mentioned properties of these dipoles. In the case
of biaxial WDD, the main contribution comes from the low-angle
scattering processes in view of the long-range character of the
perturbation energy in Eq.(2). On the contrary, for uniaxial WDD
the large-angle scattering dominates since they are strongly
screened systems (see Fig.2). In particular, at $2L\sim 10^3\div
10^{4}$\AA, $n_{i}\sim10^{9}$ cm$^{-2}$, and $\nu=0.01$ one
obtains $\rho_{bi} \sim 10^{-12}\div 10^{-11}$ $\Omega$ cm while
$\rho_{uni} \sim 10^{-9}\div 10^{-7}$ $\Omega$ cm. Notice that the
value of $\rho_{uni}$ agrees with that in the case of edge
dislocations of a similar density~\cite{rider,blewitt}.

It should be noted that the experiments show an increase of the
residual resistivity of nanocrystalline metals with $L$
decreasing~\cite{gusev}. This can be explained within the
above-proposed model due to a direct $L$-dependence of $n_i$.
Indeed, it is reasonable to assume that $n_i\sim L^{-p}$ with
$1<p\leq 2$ (usually $p=2$). Thus, in accordance with Eq.(12)
$\rho_{bi}$ increases with $L$ decreasing. It is interesting to
note that in this case $\rho_{uni}$ decreases with $L$ (except
$p=2$), as is seen from Eq.(13). Let us reiterate that only the
biaxial WDD with nonskew axes of rotations (BWDD) simulates the
grain boundary.

\subsection{Phonon scattering: low-temperature heat transport
in polycrystals}

It is clear that phonon scattering due to WDD will also affect the
low-temperature thermal conductivity, $\kappa$. We start from the
well known kinetic formula
\begin{equation}
\kappa=\frac{1}{3}\int_{0}^{\omega_{D}}C(\omega,T)v_{s}l(\omega,T)d\omega,
\end{equation}
where $C(\omega,T)d\omega$ is the specific heat contributed by acoustic
phonons within
the frequency interval $d\omega$, $v_s$ is an average phonon velocity, and
$l(\omega,T)$ is the phonon mean free path. It is suggested that
$C(\omega,T)$ has standard form with quadratic in $\omega$ density of
states and the Debye cutoff $\omega_{D}$.

The effective perturbation energy caused by the strain field of
WDD is determined as~\cite{ziman,klemens} $U({\bf r})=\hbar \omega
\gamma SpE_{AB}$, where $\hbar \omega$ is the phonon energy with
wavevector ${\bf q}$, $\omega=qv_{s}$, $v_s$ is the sound velocity
(for simplicity it is assumed that three acoustic branches are
equivalent), and $\gamma$ is the Gr\"{u}neisen constant. As
previously, we suppose that incident phonons are normal to the
disclination lines, so that we deal with the two-dimensional
scattering problem. The principal difference from the case of
electron scattering is the explicit $q$-dependence of the
perturbation energy (see details in Ref.~\cite{ziman}). Namely, the
strain tensor due to WDD remains the same while the coefficient
$B$ in Eq.(2) should be replaced by $B_{ph}=\hbar
qv_{s}\gamma\nu(1-2\sigma)/(1-\sigma)$. Eqs.(4),(5) preserve their
form in this case as well.

For BWDD the phonon mean free path is found to be~\cite{krasavin1}
\begin{eqnarray}
\l^{-1}_{bi}=2D^2(2\nu L)^2n_iq\Bigl (J_0^2(2qL)+J_1^2(2qL)\nonumber \\
-\frac{1}{2qL}J_0(2qL)J_1(2qL)\Bigr ),
\end{eqnarray}
where $D=\pi\gamma(1-2\sigma)/(1-\sigma)$. For uniaxial WDD one
obtains
\begin{eqnarray}
l^{-1}_{uni}=\frac{D^2}{2}(2\nu L)^2n_iq\Bigl
(1+J_0^2(2qL)-J_1^2(2qL)-\nonumber \\
\frac{2}{qL}J_0(2qL)J_1(2qL)\Bigr ).
\end{eqnarray}
Fig.3 shows $l(\omega)$ for three types of WDD.
\begin{figure}[htb]
\epsfxsize=8.5cm \centerline{\hspace{0mm}\epsffile{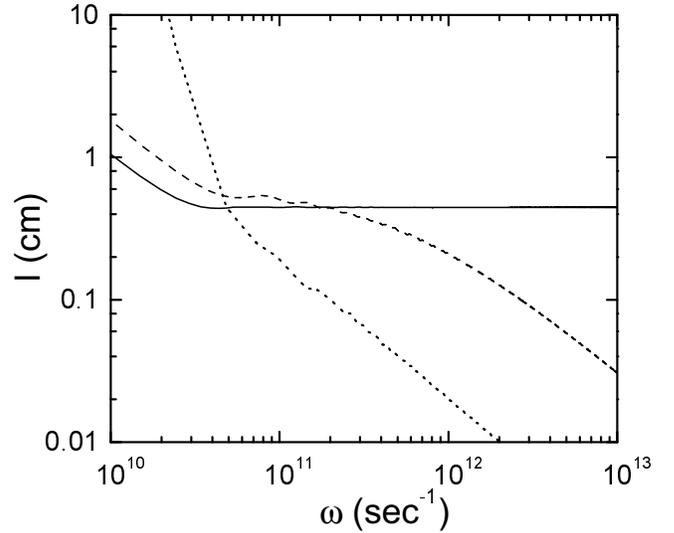}}
\vspace{2mm}
\caption{ Phonon mean free path $\l(\omega)$ at
$2q_DL=6\times 10^3$ for $\Delta_l=0$ (solid line), $\Delta_l=2L$
(dotted line), and $\Delta_l=0.5L$ (dashed line). The parameter
set used is: $L=1.35\times 10^{-5}cm$, $\nu=0.023$, $K=2.6$,
$n_i=1.8\times 10^7cm^{-2}$, and $v_s=4.8\times 10^5cm/sec$. }
\label{fig3}
\end{figure}
 We have used a typical for polycrystals size of the grain boundary $2L=2700$\AA.
As is seen, three curves behave differently. At low frequencies
the scattering by uniaxial WDD resembles that by a point impurity.
Namely, it strongly depends on $\omega$, $l_{uni}\sim \omega
^{-5}$, thus once again confirming a view of uniaxial WDD as a
strongly screened system. At high frequencies uniaxial dipoles
scatter phonons like dislocations with $l_{uni}\sim \omega ^{-1}$.
It is interesting that the same $\omega^{-1}$-dependence appears
for arbitrary biaxial dipoles both at low and high frequencies.
What is more important, there is the only type of biaxial dipoles,
BWDD, which shows the unique behavior with $l_{bi}\rightarrow
const$ as $\omega$ increases (see Fig.3). It was
found~\cite{krasavin1} that the change in behavior of $l_{bi}$
occurs at $2qL\sim 1$ or, equivalently, at $\omega^{*} \sim
v_s/2L$. It should be emphasized that this intriguing result
provides the basis for the following important speculations.
Notice that some visible irregularities in Fig.3 came from rapid
oscillations of the Bessel functions near the characteristic
frequency $\omega^{*}$.\\
The BWDD-based model was successfully used for description of the
phonon transport in polycrystals~\cite{krasavin1}. For this
purpose, Eq.(14) should be integrated with the phonon mean free
path from Eq.(15). The result is shown in Fig.4 together with the
experimental data for LiF~\cite{anderson}. As is seen, the
predictions of the theory are in good agreement with the
experimental results. It is interesting to note that the behavior
of $\kappa(T)$ in our model is governed mainly by $D$, $2\nu L$
and $n_{i}$ in Eq.(15), that is by parameters which characterize a
microstructure of polycrystals.\\
In accordance with the above analysis, the thermal conductivity exhibits a crossover from a
$\kappa\sim T^{2}$ to  $\kappa\sim T^{3}$ at $T^{'}\sim 0.1$K for
a chosen set of model parameters. It should be emphasized that
such behavior of $\kappa (T)$ is specific to BWDD (which simulates
a finite wall of edge dislocations). For example, for the uniaxial
dipoles one obtains that $\kappa\sim T^{-2}$ at low temperatures
and $\kappa\sim T^{-1}$ for $T\rightarrow \Theta_{D}$, where
$\Theta_D$ is the Debye temperature.
\vskip 0.92cm

\begin{figure}[htb]
\epsfxsize=8.5cm \centerline{\hspace{-2.5mm} \epsffile{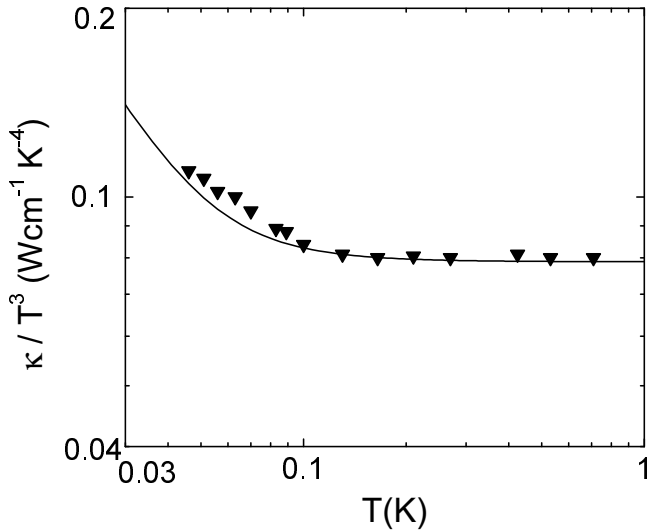}}
\vspace{2mm}
 \caption{ Reduced thermal conductivity due to biaxial
WDD scattering, $\kappa\times T^{-3}$ versus temperature $T$,
calculated according to (14) with the same parameter set as in
Fig.3. Measured points for the boundary-limited thermal
conductivity in LiF (from Ref.~\cite{anderson}) are indicated by
triangles. }
 \label{fig4}
\end{figure}

\section{BWDD in dielectric glasses: Thermal Conductivity}

There are two important consequences of the previous section.
First, it was found that the grain-boundary-induced contribution
to the thermal conductivity behaves like $T^2$ at very low
temperatures. As is well known~\cite{zeller}, this behavior is
peculiar to dielectric glasses, where $\kappa\sim T^2$ for $T<1$K.
Second, the critical temperature $T^{'}$ depends considerably on a
size of the grain. This follows from the condition $2qL\sim 1$
which can be rewritten as $T\approx \hbar v_s/2Lk_B$, where
$k_B$ is the Boltzman constant (see details in
Ref.~\cite{krasavin1}). It particular, one obtains $T\sim 1$K
at $2L\sim 20$\AA. Hence a $T^2$ dependence of $\kappa$ can be
extended up to 1K for materials consisting of microclusters with
average diameters of order $15\div 30$\AA. It is intriguing that
exactly the same values are expected within the cluster model
proposed for dielectric glasses~\cite{j1phillips}.

This finding have stimulated a detailed study of the problem. It
was shown in Ref.~\cite{krasavin2} that the experimental data for
the thermal conductivity in vitreous silica (a-SiO$_2$) can be
explained by a combination of two scattering processes. The first
one comes from the sound waves  scattering due to BWDD while the
second one is the known Rayleigh type scattering which appear due
to the local variations in structure. In this section, we present
the model and extend an analysis to other glasses.

In accordance with our scenario the mean free path is determined
to be
\begin{equation}
l(\omega)=(l_{bi}^{-1}+l_{struc}^{-1})^{-1},
\end{equation}
where $l_{bi}$ comes from Eq.(15), and $l_{struc}$ should be taken
in the most general form describing the Rayleigh scattering over
the complete frequency range. The interpolation formula reads
(see, e.g., Ref.~\cite{jones1})
\begin{equation}
l_{struc}=Y^{-1}\ (\frac{\hbar\omega}{k_{B}}\ )^{-4}+l_{0},
\end{equation}
where $Y$ is a constant which has been considered as a fitting
parameter, and $l_0$ appears as the high-frequency limit. Fig.5
shows $l_{bi}$, $l_{struc}$ and $l(\omega,T)$ with the model
parameters for a-SiO$_{2}$ taken from Table I.

\begin{figure}[htb]
\epsfxsize=8.5cm \centerline{\hspace{0mm} \epsffile{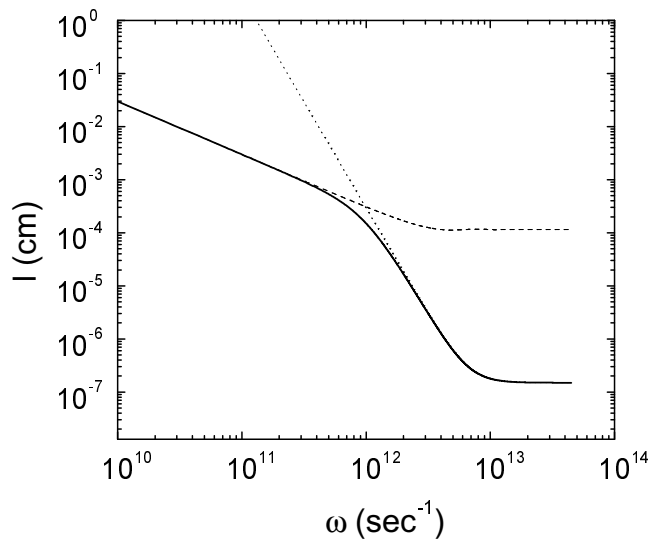}}
\vspace{2mm}
 \caption{ Phonon mean free paths $l_{bi}$ (dashed
line), $l_{struc}$ (dotted line) and $l(\omega)$ (solid line) as
functions of frequency. The parameter set for a-SiO$_2$ is shown
in Table I, K=2.6.}
\label{fig5}
\end{figure}

One can see that $l(\omega)$ has a form typical for glassy
materials. At low frequencies, $\omega<10^{12}$ sec$^{-1}$,
$l(\omega)\sim \omega^{-1}$, and the main contribution is due to
the BWDD-induced scattering. In the intermediate region both
scattering processes are involved while at high frequencies the
Rayleigh scattering becomes dominant.  Notice that the region
$10^{12}$sec$^{-1}$$<\omega<10^{13}$sec$^{-1}$ is responsible for
the plateau in the thermal conductivity. We have
found~\cite{krasavin2} that the size of this region decreases with
$2L$ and/or $l_{0}$ increasing.

It is interesting to note that Eq.(17) supports the empirical
relation $l/\lambda\sim 150$, with $\lambda$ being the wavelength
of a phonon, which holds for many glasses at low
temperatures~\cite{freeman}. Indeed, at low frequencies $l\sim
l_{bi}$. Spreading out Eq.(15) at $qL<<1$ one gets
\begin{equation}
\frac{l_{bi}}{\lambda} =\frac{1}{2\pi D^{2}(2\nu L)^{2}n_{i}}.
\end{equation}
This is a constant which depends on the model parameters which
characterize the structural and elastic properties of a material.
It is reasonable to assume that these parameters vary only
slightly in different amorphous dielectrics (see also Table I).

This can explain the observed constant-like behavior of
$l/\lambda$. In particular, for our choice of parameters for
a-SiO$_2$ one gets $l_{bi}/\lambda\sim 135$.

To calculate $\kappa$ with $l(\omega)$ from Eq.(17), it is
convenient to use the dimensionless form of Eq.(14)
\begin{equation}
\kappa = \frac{k_B^4T^3}{2\pi^2\hbar^3v_s^2}
\int_0^{\Theta_{D}/T}x^4e^x(e^x-1)^{-2}l(x)dx,
\end{equation}
where $x=\hbar\omega/k_{B}T$, and the specific heat capacity is
chosen in the standard Debye form. The results are shown in Fig.6
and Fig.7.

As is seen, there is a good agreement with the
experimental data over a wide temperature range. Notice that we
did not use any special fitting programs to get the best fit.
Instead, we have fixed the parameters related to BWDD: a dipole
separation $2L=20$\AA, the density of defects $n_i=2\times
10^{11}$cm$^{-2}$ and the Frank index $\nu=0.1$ (except those for
PS, see Table I), and tried to bring the parameters for Rayleigh
scattering close to those in Ref.~\cite{graebner}. In our opinion,
this provides better insight into the essence of the proposed
model. Let us stress once again that a characteristic length
$20$\AA \ corresponds to the expected average size of clusters
suggested in glasses~\cite{phill,j1phillips}.
\vskip 0.7cm
\begin{figure}[htb]
\epsfxsize=8.5cm
\centerline{\hspace{0mm} \epsffile{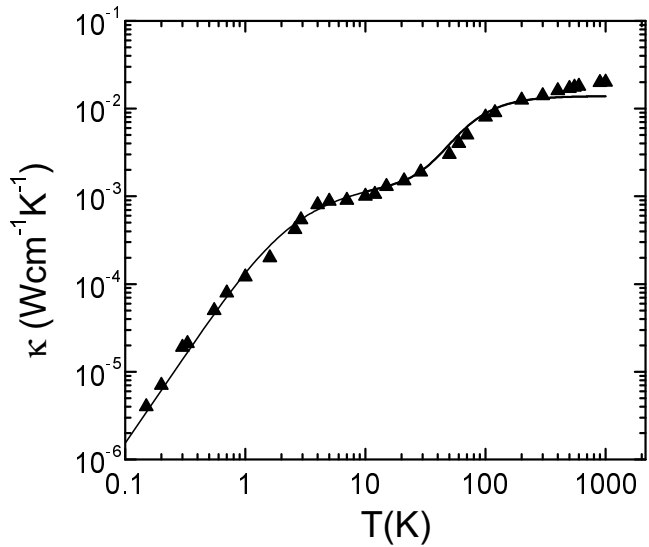}}
\vspace{3mm}
\caption{ Thermal conductivity vs temperature for
a-SiO$_2$ calculated according to Eq.(20) with $l(\omega)$ from
Eq.(17) with a set of parameters from Table I. Experimental data
from Ref.~\cite{zeller} are indicated by triangles. }
\label{fig6}
\end{figure}

\begin{figure}[htb]
\epsfxsize=8.5cm \centerline{\hspace{0mm}\epsffile{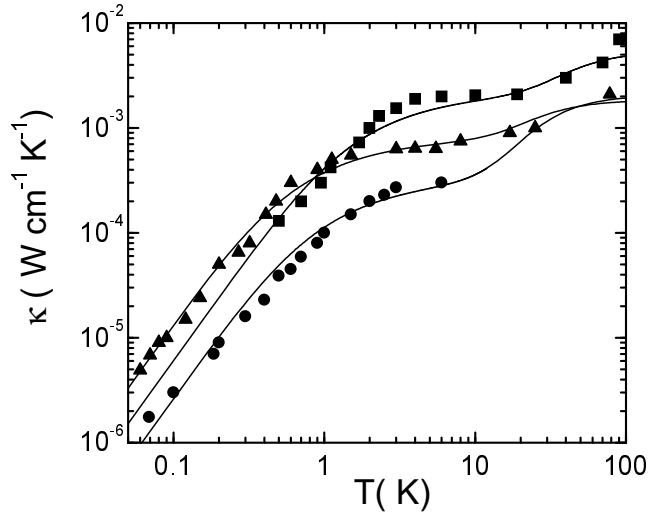}}
\vspace{3mm}
\caption{Thermal conductivity vs temperature for
a-GeO$_{2}$ (squares); a-Se (triangles) and polystyrene (PS)
(circles) (experimental data are taken from Ref.~\cite{zeller}).
Theoretically calculated  curves are represented by solid lines.
The set of the model parameters is given in Table I. }
\label{fig7}
\end{figure}

\section{Discussion }

The results obtained in the previous section call for an
additional discussion. First of all, the main question arises of
whether the proposed model captures the essence of the glassy
state or a good fitting of experiments is purely accidental. At
the moment we cannot provide the ultimate answer to this question.
It is interesting, however, to consider the peculiarities of our
approach in relation to another known ways of tackling the
problem.

As is well known, the very successful explanation of the
very-low-temperature ($T<1$K) behavior of amorphous dielectrics
has been done within the phenomenological TLS
model~\cite{varma,phillips}. Namely, at low temperatures the
principal scatterers of acoustic phonons in glasses were proposed
to be the tunneling states. As is known, however, this view posed
some important questions. First, the microscopic basis for the TLS
is still unclear~\cite{waphil,galperin}. Second, the quantitative
universality seen in various glasses at low temperatures (for
example, the above-mentioned relation $l/\lambda\sim 150$) has not
been explained. Namely, according to TLS model
\begin{equation}
l/\lambda\propto (\bar P)^{-1}\propto T_{g}/V_{F},
\end{equation}
where $\bar P$ is the density of TLS, $T_{g}$ is the glass
transition temperature, and $V_{F}$ is the free volume frozen into
the glass. However, the experimental data show that the above
relation does not depend explicitly on $T_{g}$ (see, e.g.,
discussion in Ref.~\cite{freeman}).

Third, the universal properties of glasses in the
intermediate-temperature range, $1<T<10$ K, were not understood
even qualitatively within the TLS
model~\cite{jones,freeman,karpov}. Some of these problems,
however, have been solved later by invoking additional to the TLS
concepts. In particular, a reasonable expression for the total
phonon mean free path that allows to describe $\kappa $ over a
wide temperature range reads (see, e.g.,
Refs.~\cite{graebner,sethna1})
\begin{equation}
l(\omega,T)=(l_{t}^{-1}+l^{-1}_{add}+l_{R}^{-1})^{-1},
\end{equation}
where $l_{t}$ and $l_R$ are due to TLS  and the Rayleigh
scattering, respectively, $l_{add}$ comes from some additional
scattering mechanisms. As a possible candidate there was
considered the phonon scattering from some kind of disorder
(clusters~\cite{graebner}, fractals~\cite{alexander}, etc.). The
modern theories involve the phonon scattering from localized
low-frequency vibrations~\cite{karpov,yu,sethna1} manifested
themselves both experimentally and by computer
simulations~\cite{buchenau,zhou}. All these successful approaches
are essentially based on the TLS picture. Unfortunately, the
nature of either TLS or the localized vibrations remains still
unclear.

Let us discuss these points in the context of our model. Our
consideration assumes the new principal scatterer in amorphous
dielectrics which has a clear physical origin. Indeed, as
indicated above, there is a direct analogy between BWDDs, finite
dislocation walls, and grain boundaries. Thus, the BWDD-based
model supports the cluster picture of amorphous
dielectrics~\cite{phill,j1phillips} where boundaries between
clusters give rise to defects such as BWDD. In this connection the
question arises: is there any sound experimental evidence for the
existence of either clusters or BWDDs in glasses? At present the
answer is no. As a possible reason one can mention a too small
average size of clusters as well as their random distribution. In
such an event, both clusters and BWDD are difficult to detect. Let
us stress that we have considered in our model the randomly
oriented dipoles (see averaging in Eq.(4)).

In metals, however, the phase with long-range orientational order
and no translational symmetry has been experimentally observed by
the x-ray scattering~\cite{shechtman,horn}. This finding confirms
the proposed point of view that supercooled liquids and metallic
glasses can be viewed as defected states, including disclinations,
of icosahedral bond orientational
order~\cite{nelson,kleman1,sadoc}. In particular, in two
dimensions liquids are regarded as hexatic fluids interrupted by
point disclinations (i.e. local points of 5- and 7-fold
symmetry)~\cite{halperin}. In accordance with their
scenario~\cite{halperin,nelson,nelson2}, there is a two-stage
pairing process: disclinations first pair to form 5-7 dipoles
regarded as dislocations which then pair at lower temperature to
form a crystalline solid. It is interesting that grain boundaries
are suggested to be linear arrangements of the form
--5-7--5-7--5-7--.

Regarding Eq.(21), we have shown in the previous section that the
WDD-based model predicts a constant for the relation $l/\lambda$
at low temperatures (see Eq.(19)). It is interesting that this
constant depends only on the model parameters which characterize
the structural properties of amorphous dielectrics. Let us stress
also that though Eq.(17) takes into account only two principal
scatterers, it allows to describe the thermal conductivity of
various dielectric glasses over a wide temperature range.

Another important question is the low-temperature specific heat of
glasses, $C_v$. As is known~\cite{zeller}, $C_v$ is characterized
by an anomalous linear temperature behavior. An explanation based
on the TLS model looks quite correct. While this problem is beyond
the scope of our paper, nevertheless, we can discuss briefly an
expected contribution to the specific heat due to BWDD. A similar
problem has been considered long time ago~\cite{granato,couchman}.
Granato analyzed the pinned-dislocation contribution to the
specific heat and found that at low temperatures~\cite{granato}
\begin{equation}
C_v = \frac{p\pi^2}{3}\frac{n_da^2}{Z}\frac{Nk_B}{\Theta_D}T,
\end{equation}
where $p=v_s\sqrt{\rho/G}$ ($G$ is the shear modulus and $\rho$
the density of a material), $n_d$ is the dislocation density, $a$
the lattice constant, $Z$ the number of atoms per unit cell, and
$N$ the number of atoms per mole. This result has been discussed
in connection with dielectric glasses in Ref.~\cite{couchman}. In
particular, it was shown that there is a satisfactory agreement
with the experimentally observed data for some glasses. As is
known~\cite{li1}, with decreasing of the dipole separation $2L$
the biaxial WDD becomes equivalent to an edge dislocation with the
Burgers vector $b=4L\tan(\Omega/2)$. Thus, the above result in
Eq.(23) for dislocations should be valid for the BWDD-based model
where $2L$ is expected to be very small.

Notice also that a possible explanation of the specific heat
behavior both at $T\leq 1$K and $1\leq T\leq 10$K has been given
within the soft atomic potential model~\cite{gurevich} as well as
within the elastic dipole model~\cite{sethna1}. These approaches
interpret the specific heat peculiarities in terms of the TLS
states and the additional quasilocal harmonic modes. As we have
already mentioned, the excess harmonic modes coexisting with sound
waves below $1$THz actually have been observed in
glasses~\cite{buchenau,zhou}. It should be recognized that the
presence of localized harmonic modes is peculiar to elastic
materials with extended defects as well. In particular, the phonon
spectrum in the presence of a dislocation was shown to possess
localized modes~\cite{kosevich,maradud}. Besides, the effect of
localized vibration modes due to linear defects on the thermal
properties was studied within the framework of the vibrating
string model of a dislocation~\cite{granato,granato1}. It is
clear, however, that this consideration should be accompanied by a
detailed study within the BWDD model. This problem invites further
investigation.

Finally, let us discuss briefly a possible universality of the
BWDD-based picture in polycrystals and glasses. In accordance with
our results, a principal feature that distinguishes polycrystals
from glasses is the size of a cluster. For example, let us
consider the alkali halides LiF, NaCl, and KBr:KCN. The first two
are polycrystals with the corresponding transport characteristics
(see Sec.II) while KBr:KCN is an example of the orientational
glass~\cite{yoreo}. Within our scenario this markedly different
behavior can be explained by the lower average size of a cluster
in KBr:KCN. As we have shown above, for this reason alone the
calculated crossover temperature can be moved from very low
temperatures to 1K. This means that, e.g., at $2L\sim 1000$\AA\
the $T^3$-dependence of $\kappa$ will appear above 0.1K. In
addition, one can expect that for such large clusters the Rayleigh
scattering will be suppressed while the umklapp processes become
of importance at high temperatures. As a result, the behavior of
$\kappa$ will be typical for polycrystals. On the other hand, for
small clusters the $T^{2}$ dependence of $\kappa $ goes up to 1K,
then the Rayleigh scattering becomes important. As is
known~\cite{ziman}, there are no umklapp processes for randomly
distributed small clusters. Thus, the proposed within our model
scatterers (BWDD and the Rayleigh-type) will be of importance even
at high temperatures. It is interesting to note that while the
Rayleigh scattering becomes dominant at high temperatures, the
proper behavior of $\kappa$ can be obtained only providing that
the BWDD-induced phonon mean free path tends to a constant.

\vskip 0.5cm
\acknowledgments

This work has been supported by the Russian Foundation for Basic
Research under grant No. 97-02-16623.

\appendix
\setcounter{equation}{0}
\setcounter{section}{1}
\begin{center}
{\bf APPENDIX A}
\end{center}
Let us find the exact expression for the perturbation energy in
Eq.(1). The WDD-induced strains $E_{ij}$ can be found by using of
the Hooke's law
\begin{equation}
E_{ij}=\frac{1}{2\mu (1+\sigma )}\Bigl [(1+\sigma )\sigma^{d}_{ij}-
\sigma \sigma^{d}_{ll}\delta_{ij} \Bigr],
\end{equation}
where $\sigma^{d}_{ij}$ are the stresses due to the WDD; $\mu$ and
$\sigma$ are the shear modulus and the Poisson constant,
respectively.

For the chosen in Sec.II geometry (see Fig.1) the WDD-induced
stresses $\sigma^d_{ij}$ are~\cite{deWit}
$$
\sigma^{d}_{xx}=\frac{\mu \Omega}{2\pi(1-\sigma)}\Bigl
[\frac{1}{2}
\ln\frac{(x+L)^2+y^2}{(x-L)^2+y^2}+\frac{y^2}{(x+L)^2+y^2}
$$
$$
-\frac{y^2}{(x-L)^2+y^2}-l_{1}\frac{(x+L)\Bigl((x+L)^2-y^2)\Bigr)}{\Bigl(
(x+L)^2+y^2\Bigr)^2}
$$
\begin{equation}
+l_{2}\frac{(x-L)\Bigl((x-L)^2-y^2)\Bigr)}{\Bigl((x-L)^2+y^2\Bigr)^2} \Bigr ],
\end{equation}
$$
\sigma^{d}_{yy}=\frac{\mu \Omega}{2\pi(1-\sigma)}\Bigl
[\frac{1}{2}
\ln\frac{(x+L)^2+y^2}{(x-L)^2+y^2}+\frac{(x+L)^2}{(x+L)^2+y^2}
$$
$$
-\frac{(x-L)^2}{(x-L)^2+y^2}-l_{1}\frac{(x+L)\Bigl((x+L)^2+3y^2)\Bigr)}{\Bigl(
(x+L)^2+y^2\Bigr)^2}
$$
\begin{equation}
+l_{2}\frac{(x-L)\Bigl((x-L)^2+3y^2)\Bigr)}{\Bigl(
(x-L)^2+y^2\Bigr)^2} \Bigr ],
\end{equation}
$$
\sigma^{d}_{zz}=\frac{\sigma \mu \Omega}{\pi(1-\sigma)}\Bigl [\frac{1}{2}
\ln\frac{(x+L)^2+y^2}{(x-L)^2+y^2}- l_{1}\frac{x+L}{(x+L)^2+y^2}
$$
\begin{equation}
+l_{2}\frac{x-L}{(x-L)^2+y^2}\Bigr ].
\end{equation}
Notice that Eqs.(A2)-(A4) describe the stresses due to a finite
wall of edge dislocations at large distances. Applying
Eqs.(A2)-(A4) in Eq.(A1) one gets all the components of the strain
tensor $E_{ij}$ and, finally, Eq.(2).

\appendix
\setcounter{equation}{0}
\setcounter{section}{2}
\begin{center}
{\bf APPENDIX B}
\end{center}

The perturbation energy given by Eq.(2) takes the following form in
polar coordinates
$$
U(r,\phi)=B\Biggl
[\frac{1}{2}\ln{\frac{r^2+2rL\cos\phi+L^2}{r^2-2rL\cos\phi+L^2}}
$$
\begin{equation}
-l_{1}\frac{r\cos\phi+L}{r^2+2rL\cos\phi+L^2}
+l_{2}\frac{r\cos\phi-L}{r^2-2rL\cos\phi+L^2}\Biggr ].
\end{equation}
The matrix element in Eq.(5) with the perturbation energy from
Eq.(B1) can be calculated using the following formulas:
\begin{equation}
\sum_{k=1}^{\infty}\frac{z^{2k-1}\cos(2k-1)\phi}{2k-1}=\frac{1}{4}
\ln{\frac{1+2z\cos\phi+z^2}{1-2z\cos\phi+z^2}},\hspace*{0.2cm} z^2\leq 1,
\end{equation}
\begin{equation}
\sum_{k=0}^{\infty}z^{k}\cos k\phi=\frac{1-z\cos\phi}{1-2z\cos\phi+z^2},\hspace*{0.2cm} \left| z \right|<1
\end{equation}
Substituting Eqs.(B2) and (B3) into Eq.(B1) and integrating in
Eq.(5) one obtains
$$
U(p,\alpha )=-i\frac{4\pi BL}{pS}\Bigl
[J_{0}(pL)\cos\alpha
$$
$$
+ \sum_{k=1}^{\infty}(-1)^{k}J_{2k}(pL)\Bigl (
\frac{\cos(2k+1)\alpha}{2k+1}-\frac{\cos(2k-1)\alpha}{2k-1}
\Bigr ) \Bigr ]
$$
\begin{equation}
+i\frac{2\pi B}{pS}\Delta_{l}\Bigl [J_{0}(pL)\cos\alpha+
2\cos\alpha \sum_{k=1}^{\infty}(-1)^{k}J_{2k}(pL)\cos2k\alpha\Bigr ] ,
\end{equation}
where $\Delta_{l}=l_{1}-l_{2}$ and $J_{m}(z)$ is the Bessel
function. The first term in Eq.(B4) comes from the integration of
logarithmic function in Eq.(B1) while the second one comes from
two last terms in the r.h.s. of Eq.(B1).

We have used the following standard integrals deriving Eq.(B4)
\begin{equation}
\int\limits_{0}^{2\pi}\exp(iz\cos\phi)\cos m\phi d\phi=2\pi i^{m}J_{m}(z),
\end{equation}
\begin{equation}
\int z^{k}J_{k-1}(z)dz=z^{k}J_{k}(z).
\end{equation}
The first sum in Eq.(B4) can be simplified by  differentiation
with respect to $\alpha$. The second sum in Eq.(B4) reads
\begin{equation}
\sum_{k=1}^{\infty}(-1)^{k}\cos (k\alpha) J_{2k}(z)=\frac{1}{2}\cos[z\cos(\alpha/2)]
-\frac{1}{2}J_{0}(z)
\end{equation}

After straightforward calculations one obtains
\begin{eqnarray}
\label{eq6}
U(p, \alpha)=\frac{B}{S}\Bigl [
-\frac{4\pi i}{p^2}\sin(pL\cos\alpha) \nonumber
&+&\frac{2\pi i\Delta_l}{p}\cos\alpha\cos(pL\cos\alpha)\Bigr ].
\end{eqnarray}

To find $\Re (\theta)$ in Eq.(4) one has to make averaging of
$\left|U(p,\alpha)\right|^2$ over $\alpha$:
$$
\left|U(p)\right|^{2}=\overline{\left|<{\bf q}|U({\bf r})|{\bf
q'}>\right|^2}=\frac{1}{2\pi}
\int_{0}^{2\pi}\left|U(p,\alpha)\right|^{2}d\alpha =
$$
$$
\frac{2\pi
B^{2}}{S^2}\int_{0}^{2\pi}\Bigl(\frac{4}{p^4}\sin^2(pL\cos\alpha
)- \frac{2\Delta_l}{p^3}\cos\alpha \sin(2pL\cos\alpha)
$$
\begin{equation}
+\frac{\Delta_l^2}{p^2}\cos^2\alpha\cos^2(pL\cos\alpha)
\Bigr)d\alpha.
\end{equation}
Using
\begin{eqnarray}
&&\int_{0}^{2\pi}{\cos(z\cos\alpha)\choose{\sin(z\cos\alpha)}}\cos(n\alpha)d\alpha
=2\pi{\cos(n\pi/2)\choose{\sin(n\pi/2)}}J_{n}(z), \nonumber \\
&&
\end{eqnarray}
one finally gets
\begin{eqnarray}
\left|U(p)\right|^{2}=\frac{4\pi^2B^2}{S^2}\Biggl\{\frac{2}{p^4}\Bigl(1-J_{0}(2pL)\Bigr)
-\frac{2\Delta_{l}}{p^3}J_{1}(2pL)\nonumber \\
+\frac{\Delta^2_l}{2p^2}\Bigl(\frac{1}{2}+J_{0}(2pL)-\frac{J_{1}(2pL)}{2pL}\Bigr) \Biggr \}.
\end{eqnarray}
Substituting Eq.(B10) into Eq.(4) one obtains the effective differential
scattering radius in Eq.(6).

The exact expression for the mean free path in Eq.(3) takes the
form
\begin{equation}
\Lambda ^{-1}=\frac{n_iB^2\pi }{\hbar ^2v^2}\Biggl\{\frac{1}{2q^3}I_{1}(qL)+
\frac{\Delta ^2_l}{2q}I_2(qL)-\frac{\Delta _l}{q^2}I_{3}(qL)\Biggr\},
\end{equation}
where
$$
I_1(qL)=\int_{0}^{2\pi}\frac{d\theta
}{\sin^2(\theta/2)}(1-J_{0}(4qL\sin(\theta/2))=
$$
\begin{equation}
16q^2L^2\pi\Bigl(J^2_{0}(2qL)+J^2_{1}(2qL)\Bigr)-8qL\pi J_0(2qL)J_1(2qL),
\end{equation}

$$
I_2(qL)=\int_{0}^{2\pi}\Bigl(\frac{1}{2}+J_{0}(4qL\sin(\theta /2))
-\frac{J_1(4qL\sin(\theta /2))}{4qL\sin(\theta /2)}\Bigr){d\theta}=
$$
\begin{equation}
2\pi\Bigl(\frac{1}{2}+J^2_{0}(2qL)\Bigr)-\pi \Bigl(J^2_{0}(2qL)+J^2_{1}(2qL)\Bigr),
\end{equation}

$$
I_3(qL)=\int_{0}^{2\pi}\frac{d\theta
}{\sin(\theta/2)}J_{1}(4qL\sin(\theta/2))=
$$
\begin{equation}
4qL\pi\Bigl(J^2_{0}(2qL)+J^2_{1}(2qL)\Bigr).
\end{equation}'

The final result for the mean free path is given by Eq.(7).

\begin{table}

\caption{Parameters used for the numerical fits. Units are the
following: $v_s$ ($10^5$ cm/sec), $\Theta_D$ (K), $n_i$ ($10^{11}$
cm$^{-2}$), $2L$ ($10^{-7}$ cm), $Y$ (cm$^{-1}$K$^{-4}$), $l_{0}$
($10^{-7}$ cm).}

\label{ll}

\begin{tabular}{lccccccc}
{\small  Material} &$v_s$&$\Theta_D$&$n_i$&$2L$&$\nu$&$l_0$&$Y$\\
\tableline
a-SiO$_2$&4.1&342&2&2&0.1&1.5&1\\
a-GeO$_2$&2.6&192&2&2&0.1&0.6&2.9\\
a-Se&1.19&113&2&2&0.1&0.2&90\\
PS&1.67&123&5&2.4&0.1&0.5&80\\
\end{tabular}
\end{table}
\end{multicols}
\end{document}